# The Influence of Social User Knowledge Level and Active Communication Channel Control on Rumor Spread

*Yixuan Zhao& Rohitha Settipalli*

*Department of Geography, State University of New York at Buffalo*

## Abstract

This research examines the propagation of rumors on social networks during public health emergencies and explores strategies to effectively manage false information in cyberspace. Using a simulation model, the study analyzes the impact of factors such as communication channel control, government intervention, and individual personalities on the spread of rumors. The results suggest that enhancing netizens' knowledge and capacity to recognize and resist rumors, developing rumor-debunking platforms, and promoting a "clear" ecology of network information content are effective strategies for controlling false information in cyberspace. However, the complexity and scale of actual networks present challenges to the development of a comprehensive cyberspace governance system. The findings offer practical guidelines for improving the effectiveness of governance in managing the spread of rumors on social networks.

**Introduction**

The growth of the internet has led to the pervasive spread of rumors[1]. Individuals spread rumors for various reasons, including seeking attention, defaming others or entities, gaining momentum for an idea or cause, diverting attention, or inducing panic[2]. The rapid expansion of communication tools and the acceleration of informationalization have further enabled the propagation of rumors, resulting in their wider and faster dissemination[3].

One of the biggest challenges in the field of public health is the potential for misinformation spread on social media networks to endanger societal stability and public opinion during times of public health emergencies[4]. The publication of official notices of crises often lags due to the ambiguity and lack of information surrounding them, thus uncertainty impacts how rumors spread on the Internet[5]. Currently, the likelihood of Internet rumors spreading, the ambiguity of events, the risk to public opinion, and social impact all increase in direct proportion to the questionable nature of evidence for Internet rumors[6].

---

[1] Anderson, M., & Jiang, J. (2018). Teens, social media & technology 2018. Pew Research Center, 31(2018), 1673-1689.
[2] Festinger, L., Cartwright, D., Barber, K., Fleischl, J., Gottsdanker, J., Keysen, A., & Leavitt, G. (1948). A study of a rumor: Its origin and spread. *Human Relations*, *1*(4), 464-486.
[3] Shah, D., & Zaman, T. (2011). Rumors in a network: Who's the culprit?. IEEE Transactions on information theory, 57(8), 5163-5181.
[4] Glik, D. C. (2007). Risk communication for public health emergencies. Annual review of public health, 28, 33.

The fear of Internet users and the unpredictability of information pertaining to emergencies can eventually transform public health crises into information crises in the context of information dissemination on social networks. Due to a lack of information, the public is easily swayed by the spread of rumors on the Internet, resulting in inaccurate scientific understanding and social turmoil[7]. Therefore, understanding how rumors propagate has practical applications, aiding the government in formulating efficient management strategies to safeguard public safety, social stability, and growth.

Our research examines how critical information can be managed in society, including subsequent inquiries regarding consolidating information review, raising the general population's level of education, and government intervention through propaganda or other techniques. We investigate how people's personalities, communication channel control, and governmental interference affect rumor-spreading incidents. Using the Sir-based Dynamic Model of the Analogic Construction System, we simulate the rumor-spreading process. To achieve the goal of cyberspace governance, we aim to turn the ambiguous information situation of public health emergencies into an information disclosure situation through efficient network rumor-refuting strategies. This will calm netizens' anxiety and prevent internet rumors from spreading.

**Dynamic Hypothesis**

The conventional approach to rumor dissemination in social networks considers the creation of original information as the start of the rumor event. Over time, the original information is disseminated, undergoes distortion, and becomes rumor information, initiating the rumor-spreading process[8][9]. In the digital age, the SIR model is commonly used to study rumor dissemination. After a specific amount of time, the rumor information is generated and the rumor-debunking process begins[10]. Currently, rumor information and refuting information coexist in the network, resulting in a competitive communication relationship between the two. The replacement process demonstrates how users change from a state of rumor-spreading to rumor-refuting in their dissemination state[11]. This is a one-way replacement relationship, where contact with refuting information will replace the rumor node with refuting information. The rumor event concludes when both pieces of information fade away[12]. This study focuses on the competitive dissemination process of rumor information and the impact of the recently added one-way permutation process on this competition.

---

[5] Zhang, L., Chen, K., Jiang, H., & Zhao, J. (2020). How the health rumor misleads people's perception in a public health emergency: lessons from a purchase craze during the COVID-19 outbreak in China. International journal of environmental research and public health, 17(19), 7213.

[6] Sun, R., An, L., Li, G., & Yu, C. (2022). Predicting social media rumours in the context of public health emergencies. Journal of Information Science, 01655515221137879.

[7] Rocha, Y. M., de Moura, G. A., Desidério, G. A., de Oliveira, C. H., Lourenço, F. D., & de Figueiredo Nicolete, L. D. (2021). The impact of fake news on social media and its influence on health during the COVID-19 pandemic: A systematic review. Journal of Public Health, 1-10.

The figure1 below illustrates the process of rumor dissemination. This schematic was developed based on prior research.

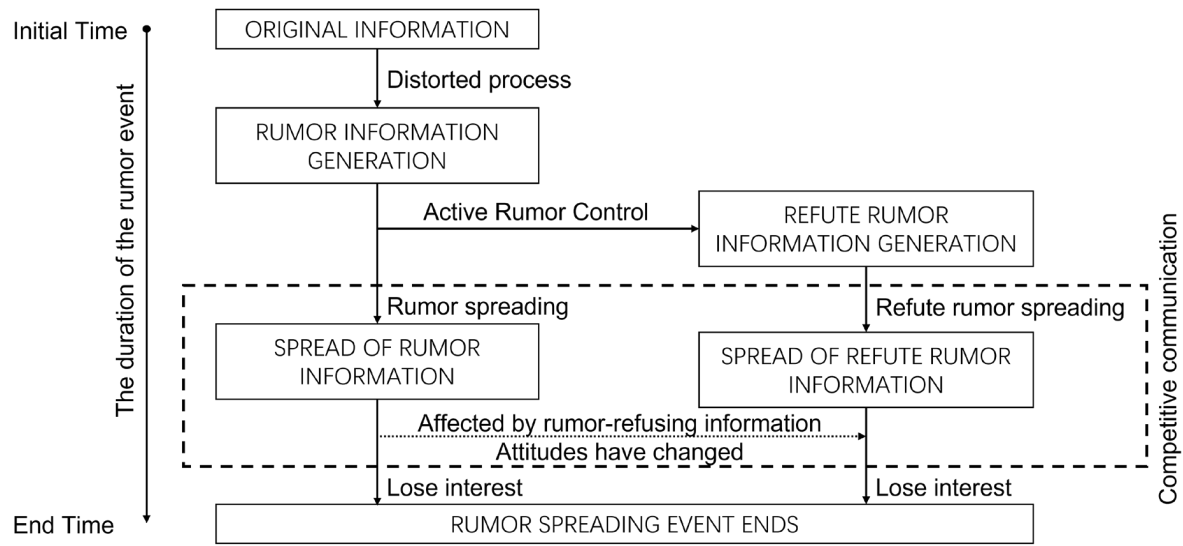

Figure 1. Process of rumor dissemination

---

[8] Zhao, L., Cui, H., Qiu, X., Wang, X., & Wang, J. (2013). SIR rumor spreading model in the new media age. *Physica A: Statistical Mechanics and its Applications*, *392*(4), 995-1003.
[9] Huang, H., Chen, Y., & Ma, Y. (2021). Modeling the competitive diffusions of rumor and knowledge and the impacts on epidemic spreading. *Applied mathematics and computation*, *388*, 125536.
[10] Yang, L., Li, Z., & Giua, A. (2019, July). Rumor containment by spreading correct information in social networks. In *2019 American Control Conference (ACC)* (pp. 5608-5613). IEEE.
[11] Lim, Y., Ozdaglar, A., & Teytelboym, A. (2017). Competitive rumor spread in social networks. *ACM SIGMETRICS Performance Evaluation Review*, *44*(3), 7-14.
[12] Wang, Z., Liang, J., Nie, H., & Zhao, H. (2020). A 3SI3R model for the propagation of two rumors with mutual promotion. *Advances in Difference Equations*, *2020*(1), 1-19.

**Method**

As rumor dissemination and disease dissemination share similarities, epidemic models are often utilized as a basis for studying rumor dissemination. The Susceptible-Infected-Recovered (SIR) model is the most widely recognized one, and subsequent studies on network rumor models have mostly used it as a base model and optimized it[13]. In order to comprehend the broad mechanism of rumor event dissemination and its impact on the general public, particularly during the COVID-19 pandemic, we utilized AnyLogic to create an SIR-based epidemic model. Our aim is to investigate the competitive dissemination process of the model and develop countermeasures that benefit the control of false information in cyberspace. Therefore, we conducted research using the model simulation approach, which is based on the model's dynamics and relationships, and involved the appropriate setup of model parameters to model and analyze the main influences on the competitive transmission process of rumor information and factual information.

**Model design**

Rumor information often emerges after factual information. In the competitive communication process, there is a difference between the initial node number IA of spreading rumor information and the initial node number IB of spreading rumor information after the fermentation time difference. When determining the initial value of the simulation time difference, IA>IB is taken into account. Our analysis shows that the dissemination rate of rumor information order $b_2$ and the replacement rate o are related to netizens' internet literacy in cyberspace. While these three parameters remain stable over a short period of time, the quality of netizens changes dynamically over a longer period. Therefore, in our model, $b_1$, $b_2$, and o are all adjustable variables.

Information service platforms can use algorithms and other technical methods to determine the user's recommended content. However, effective contact weights $w_1$, $w_2$, and $w_3$ in the rumor spreading, rumor dispelling, and replacement processes are all affected by the subjective factors of the user's browsing behavior. As the intervention changes the weight of the user's access to relevant information, w1, w2, and w3 are also adjustable variables in our model.

---

[13] Zhao, L., Cui, H., Qiu, X., Wang, X., & Wang, J. (2013). SIR rumor spreading model in the new media age. *Physica A: Statistical Mechanics and its Applications*, *392*(4), 995-1003.

On the other hand, the variables r1 and r2 in our model cannot be adjusted since the factors affecting the rumor information fade-out rate r1 and the rumor-refusing information fade-out rate r2 are more complicated and impossible to manage through external measurements. We use Anylogic, a Java-based system modeling and simulation tool, to simulate the competitive spread process of rumor information and rumor information based on the preceding system dynamics differential equations. The simulation model is illustrated in Figure 2. It includes four state nodes and associated parameters, and all of them follow the dynamic relation. We use AnyLogic as a modeling tool to simulate this process as shown in Figure 3.

($I_B$) Network nodes are in the state of having received refutation information and actively spreading refutation information
($I_A$) The network node is in the state of receiving rumor information and actively spreading rumor information
($S$) Network nodes are in the state of not spreading rumor information and refuting rumor information
($R$) The network nodes are in the state of fading out of the rumor event
($o$) The replacement rate of IA state to IB state during propagation
($b_2$) Dissemination rate of rumor−refuting information
($b_1$) Spread rate of rumor information
($y_2$) The fading rate of refuting information
($y_1$) Extinction rate of rumor information
($w$) Number of Contacts

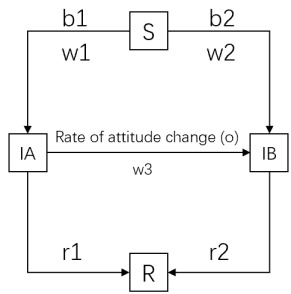

$$S(t) + I_A(t) + I_B(t) + R(t) = N$$

$$\begin{cases} \dfrac{d(S)}{d(t)} = -\dfrac{w_1}{N} I_A(t)S(t) - \dfrac{w_2}{N} I_A(t)S(t) \\ \dfrac{d(I)}{d(t)} = b_1 \dfrac{w_1}{N} I_A(t)S(t) - o \dfrac{w_3}{N} I_A(t)I_B(t) - r_1 I_A(t) \\ \dfrac{d(I)}{d(t)} = b_2 \dfrac{w_2}{N} I_B(t)S(t) - o \dfrac{w_3}{N} I_A(t)I_B(t) - r_2 I_B(t) \\ \dfrac{d(R)}{d(t)} = r_1 I_A(t) + r_2 I_B(t) \end{cases}$$

Figure 2. The Logical Formula of Rumor Propagation System Dynamics

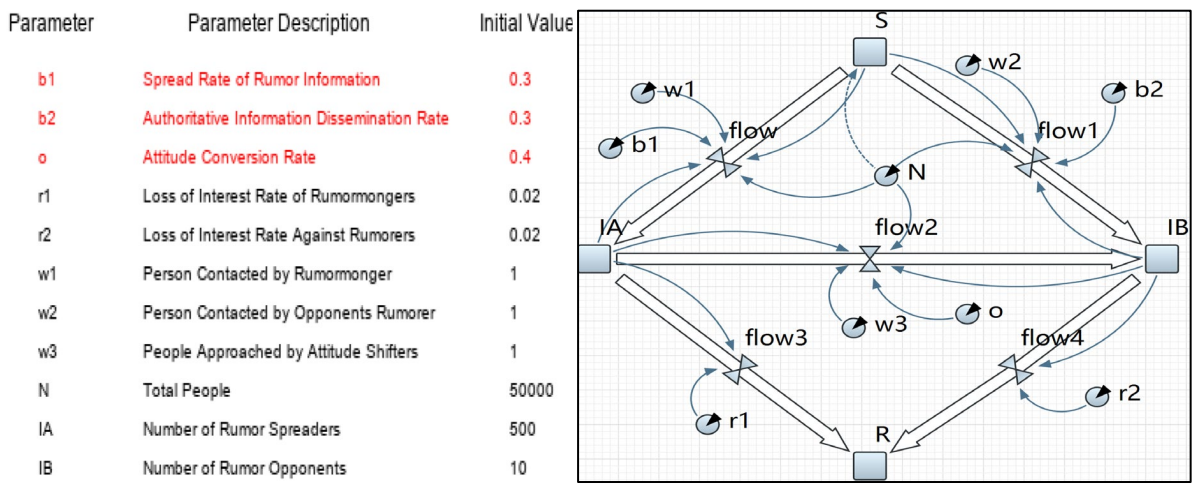

Figure 3. AnyLogic model logic implementation

## Results

**I.** *Initial value conditions*

The simulation results of the model for the initial value conditions are presented in the analysis. As IA(t) and IB(t) evolve at different rates over time, the peak of rumor information dissemination in the initial value state occurs after that of rumor information. Therefore, analyzing the changes in the number of IA rumor statuses and the number of IB rumor dispelling statuses in the simulation graph can help clarify the key regulatory factors in the process of competitive dissemination of rumor events and find potent measures for rumor governance. Crucial observations have been made.

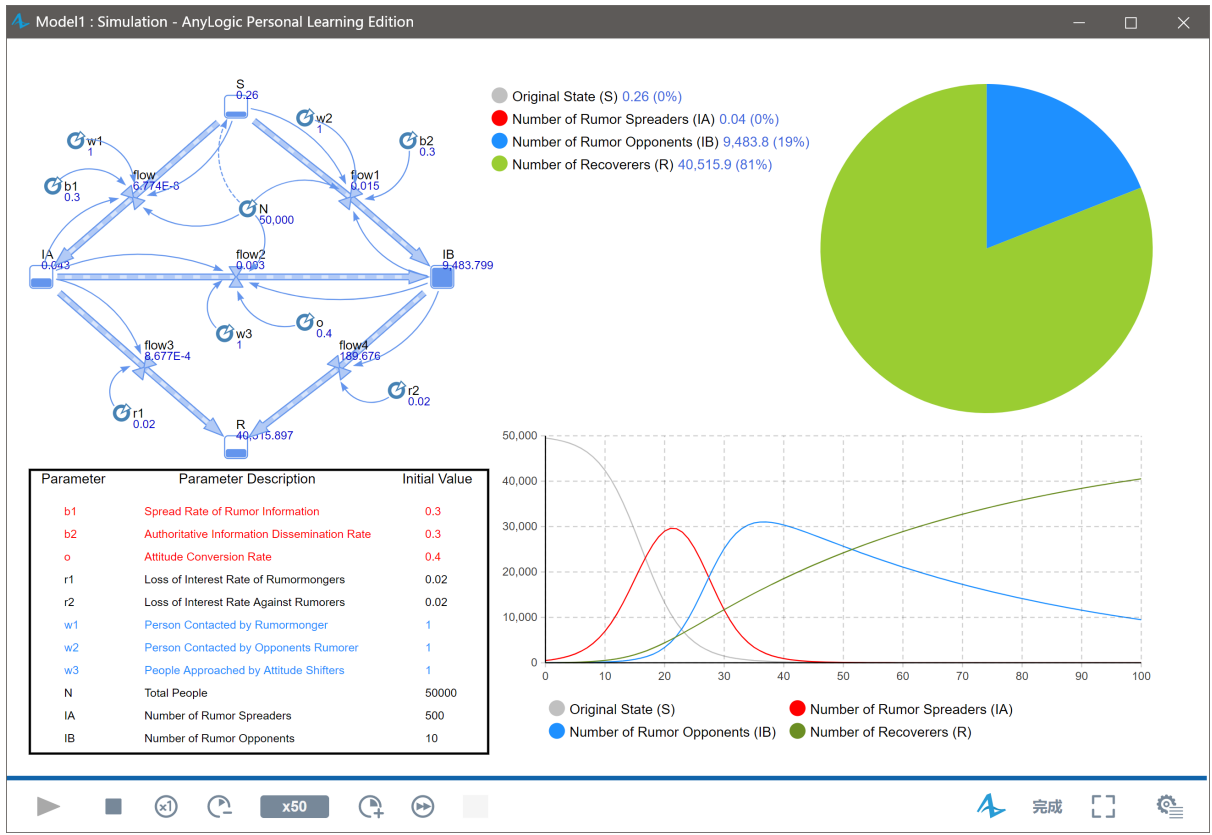

Figure 4.

*II The influence of the quality of the netizens on the spread of the rumors:*

In our study, we simulated the impact of netizen caliber on the spread of rumors. To further investigate the impact of the three parameters (b1, b2, and o) on the state number of IA spread rumors and to avoid errors caused by parameter adjustment, we conducted a sensitivity analysis on each parameter, and the direction of change was indicated by an arrow. Figures 5 and 6 show the results of these simulations for low and high netizen quality, respectively.

When the general quality of netizens is low, their ability to recognize rumors is weak, making it easier for them to participate in spreading rumor information. In this case, we set b1 > b2. As b increases, the scale of IA rumor-spreading states grows, the peak value rises, and the primary propagation period marginally extends. However, the scale of IB rumor-spreading and refuting states contracts.

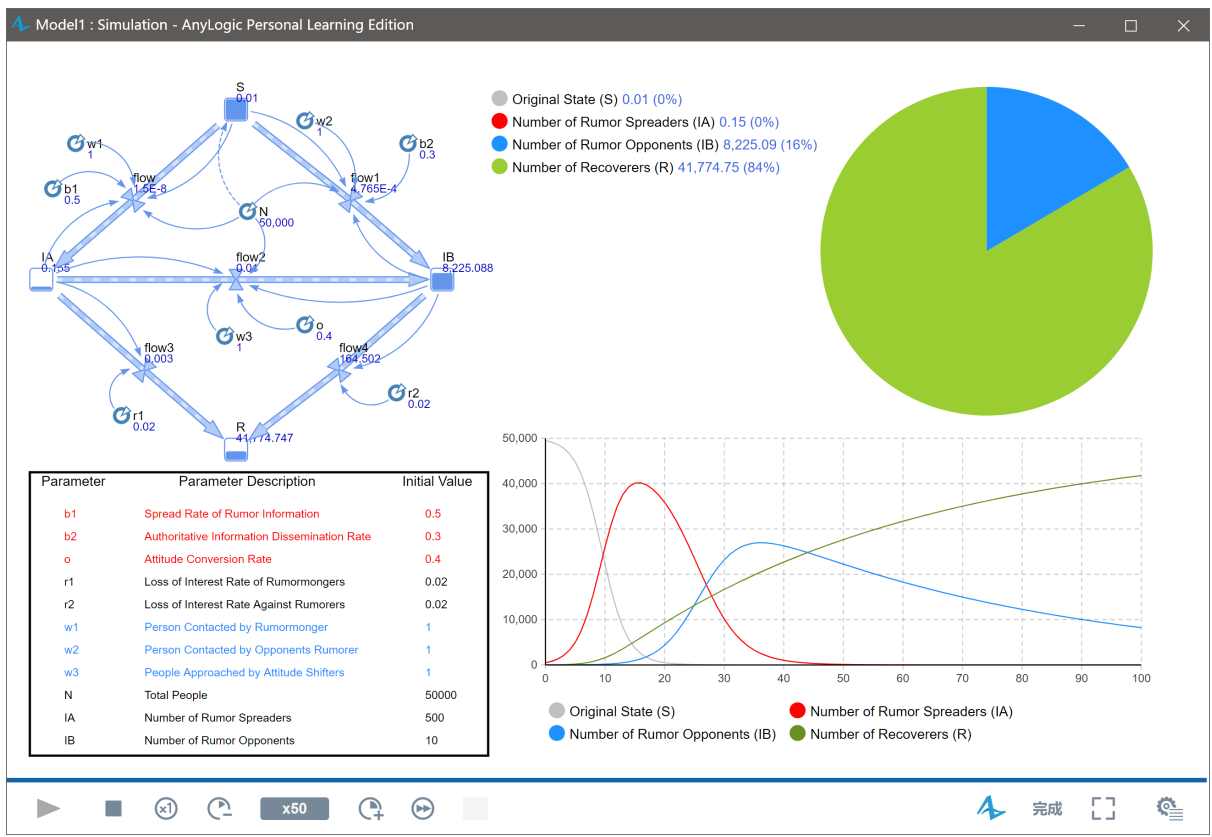

Figure 5. Simulation general quality of netizens is low (b1 > b2)

On the other hand, earlier studies suggest that the ability to recognize rumor information is strong and the desire to disseminate rumors is low when the general quality of netizens is high. To simulate this scenario, we set b1 < b2. Our investigation revealed that as the value of o rises, the scale of the state number of IA spreading rumors declines, the peak value falls, and the main duration shortens. Conversely, the scale of the state number of IB spreading rumors rises, the peak value advances, and moves sooner.

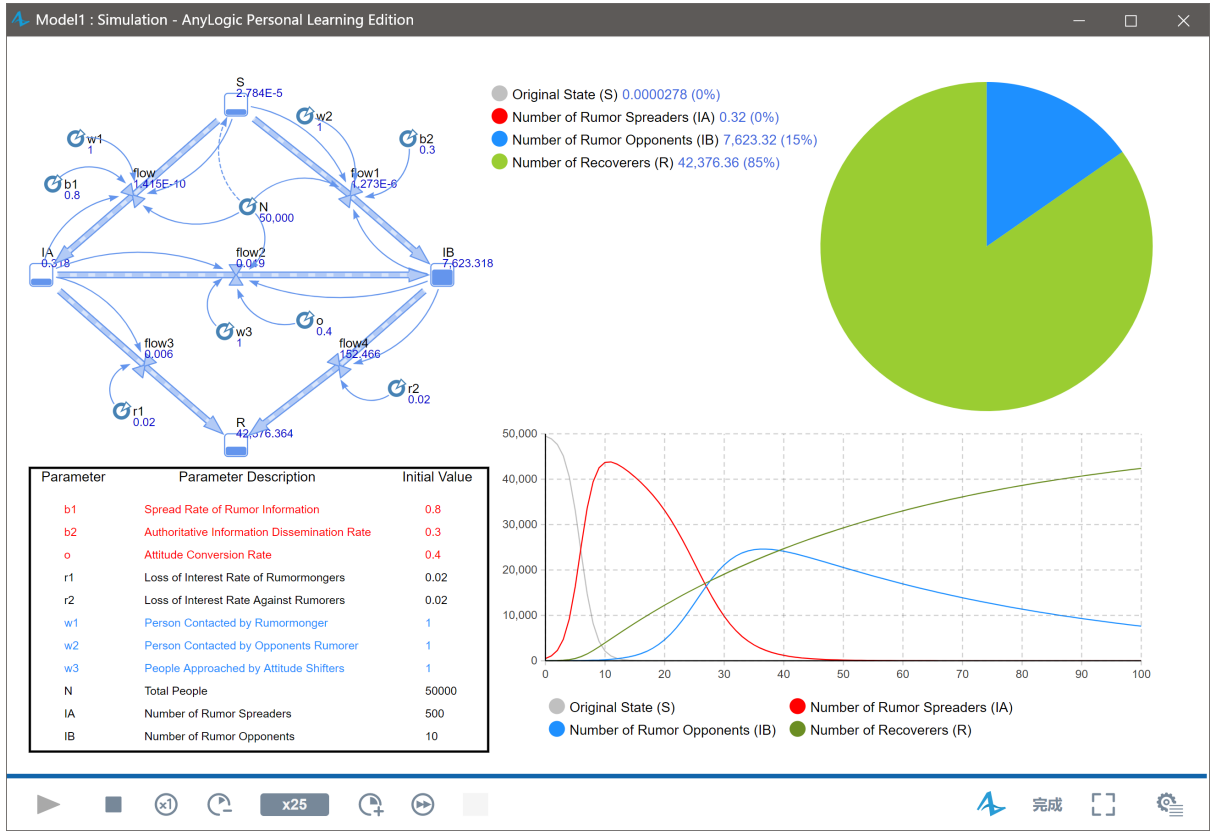

Figure 6. Simulation general quality of netizens is good (b1 < b2)

Figures 5 and 6 depict the effects of changes in the spread rate of rumor information (b1) on the number of rumor spreaders (IA), with all other parameters held constant except for b1. We found that the IA rumor status number is more sensitive to changes in the rumor propagation rate (b2) when the parameters vary between 0 and 1, with each change being 0.1 units. As more individuals give the rumor credibility, the value of IA increases.

**III** *The Influence Of Information (Contact Rate W) On The Spread Of Rumors:*

The impact of social network shutdowns on the propagation of rumors was the focus of our research. We began by simulating the spread of unchecked rumors, assuming that nodes with discrepancies between their initial IA and IB values are more susceptible to rumor information. These nodes are likely to become rumor spreaders and may even increase the spread of rumors if their contact weight in the node contact environment increases during the rumor process. To reflect this process, we varied the proportion of w1 and w2. Specifically, In figure 7 when w = 4, each rumor contacts four people per day, while in figure 8 when w = 8, each rumor spreader interacts with eight people daily.

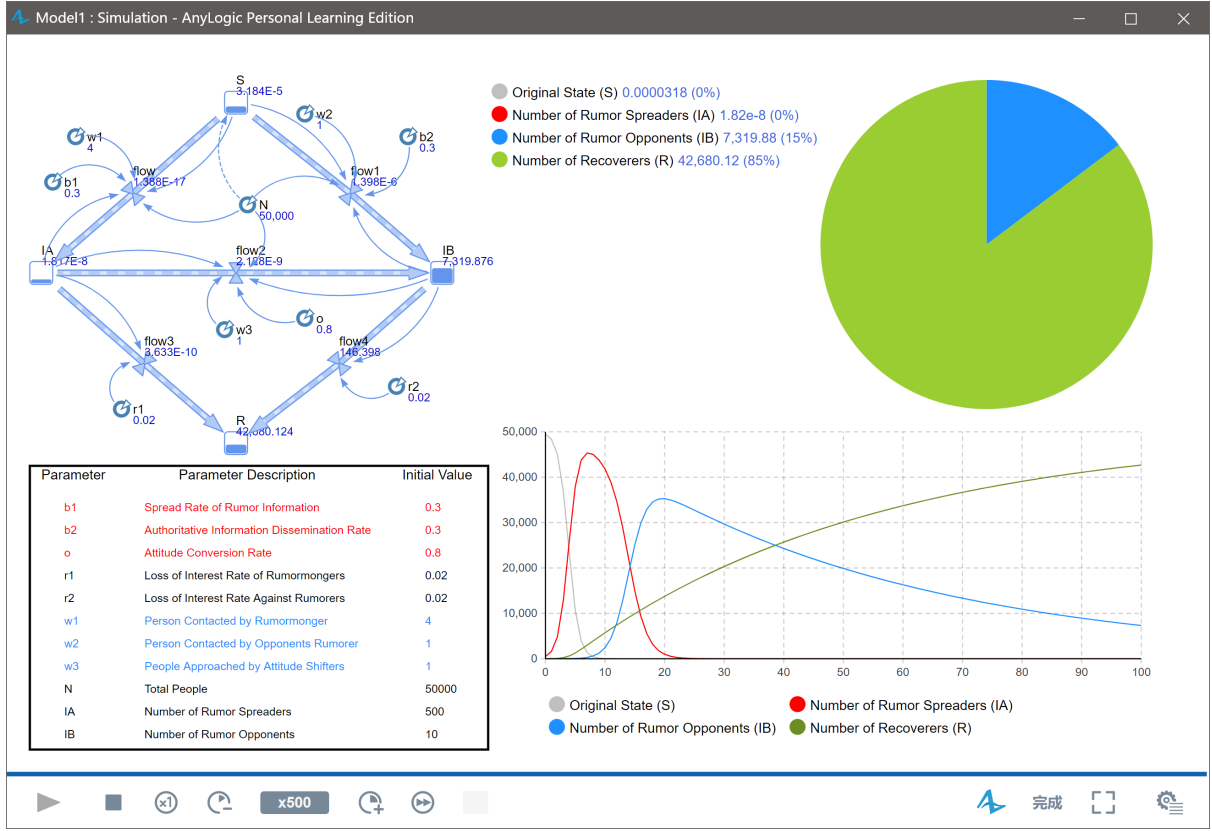

Figure 7. Simulate weak social information dissemination W = 4.

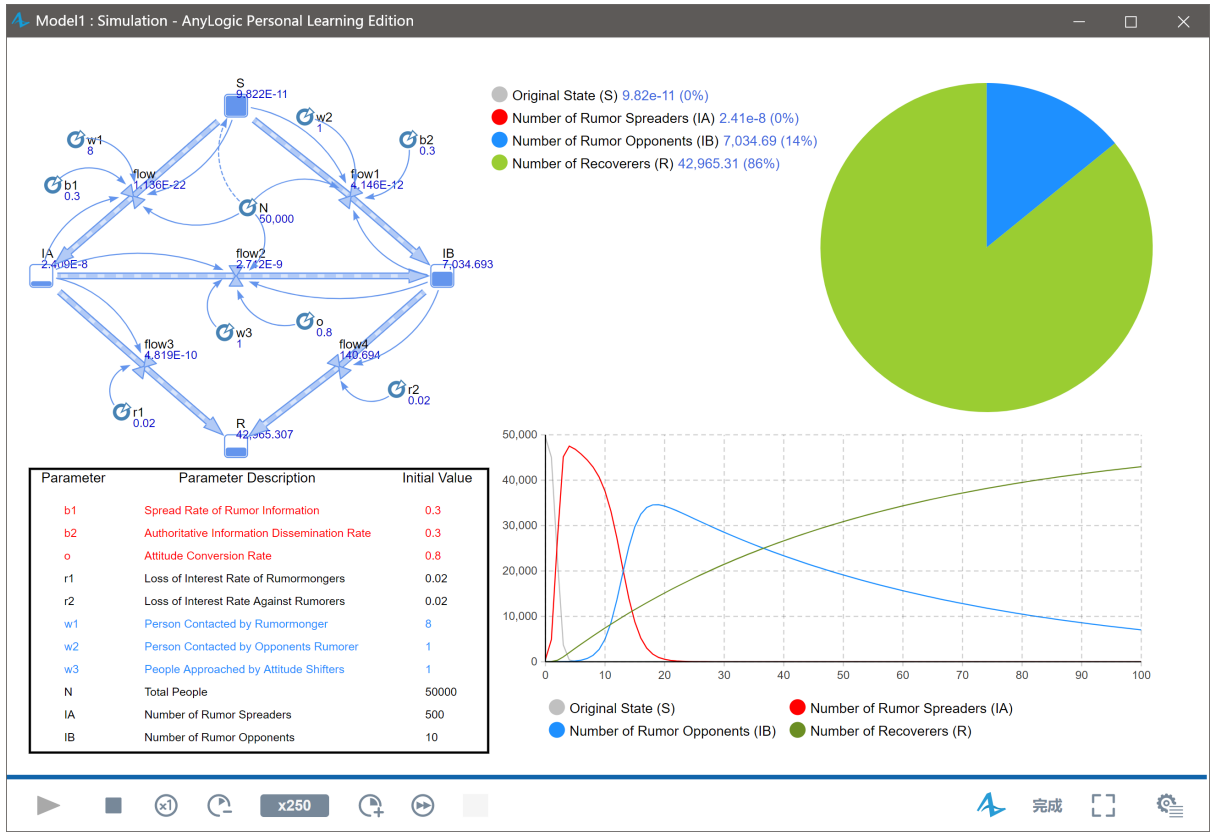

Figure 8. Simulate strong social information dissemination W = 8

Our simulations indicated that as the value of w1 increases, the number of nodes in state IA disseminating rumors decreases, while the number of nodes in state IB disseminating rumors slightly increases, resulting in a slightly expanded scale of dissemination. We then examined the effect of active rumor control on the spread of rumors. Additionally, we simulated how different rumor-busting strategies, such as widespread public broadcasting or SMS rumor-busting methods, affect the spread of rumors. Our findings revealed that during the competitive dissemination process, the rumor state node can be replaced by another rumor state node. The conversion probability of the replacement path can be improved by increasing the effective contact weight of the replacement process. Therefore, we varied w3 to observe this process. We found that as the value of w3 increases, the peak value of rumors distributed by IA falls, the time step of propagation shortens, and the scale of rumors spreads. In contrast, the scale of rumors propagated by IB rises, and the peak value increases earlier and higher.

By conducting sensitivity analyses on each variable, we demonstrate how unchecked rumors can cause unexpected societal upheaval. However, if the rumors are stopped before they spread, their impact can be minimized.

**Results**

This model's parameters, B1, B2, and O, describe the attributes of an individual's knowledge level regarding the Internet, and the experimental results of parameter adjustments are presented as follows. Based on the experiment results, raising the level of netizens' knowledge can assist in judging and discriminating between rumors, decreasing the spread rate (B1) of rumor information, and increasing the spread rate (B2) and replacement rate (O) of information that dispels rumors, thereby slowing down the spread of rumors in cyberspace. By controlling the number of nodes in the rumor state at a reasonably low level and achieving the goal of purifying the network space, the length and scale of rumor dissemination are decreased. Combining the simulation results with sensitivity analysis reveals that the rumor dissemination rate (B2) is a significant factor in the parameters related to netizens' knowledge level, and its alteration has the most visible impact on the regulation of Internet rumors.

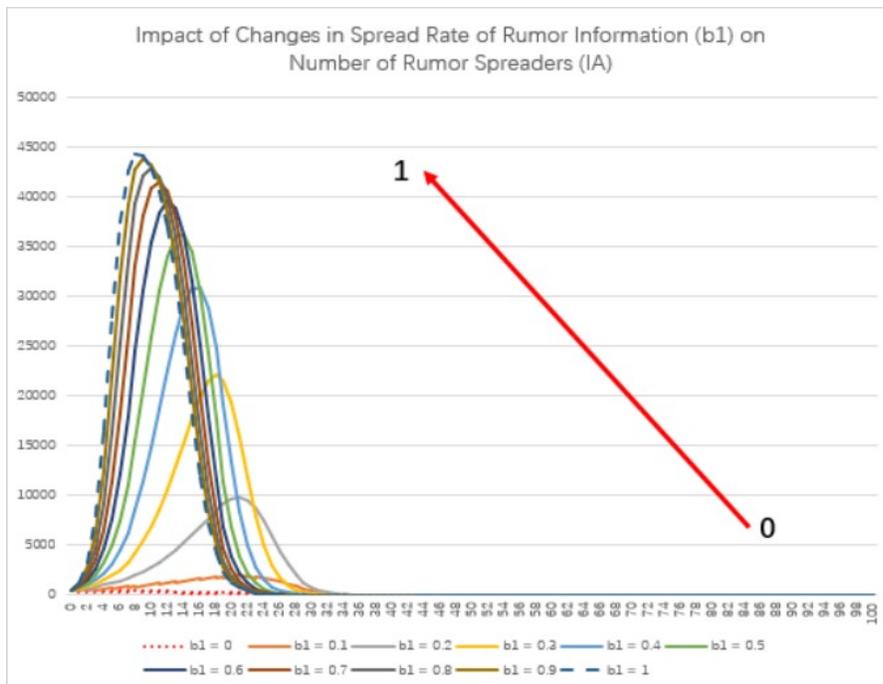

Figure 9a.

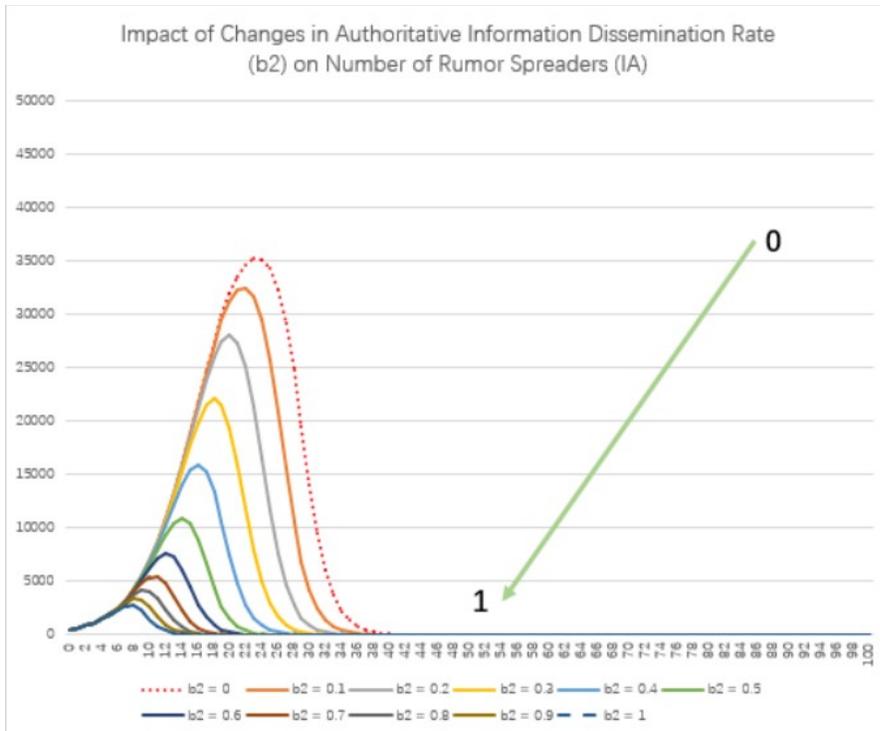

Figure 9b.

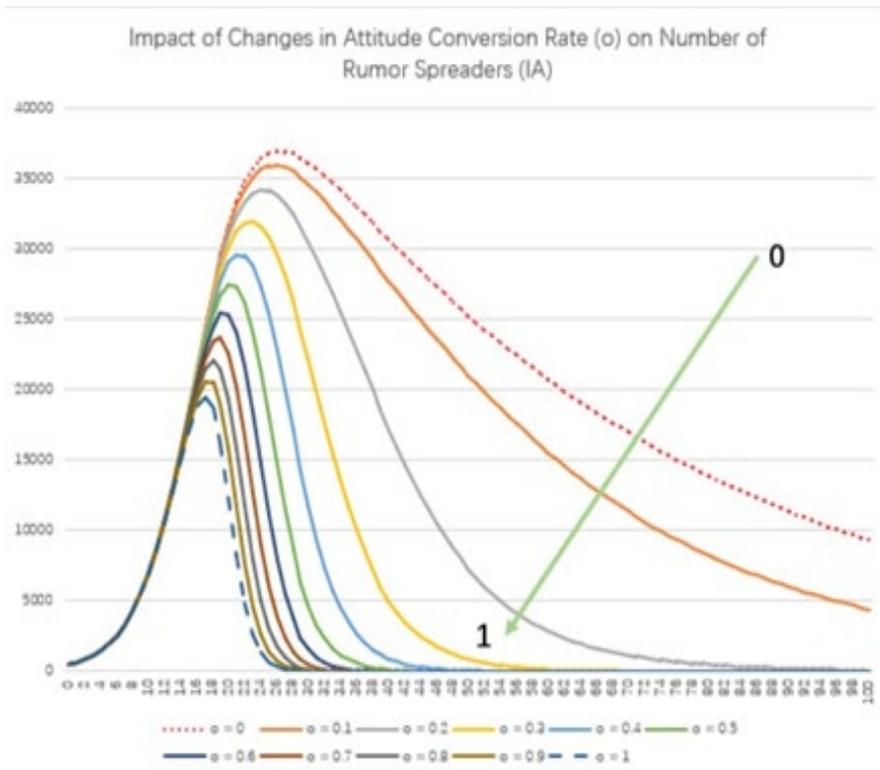

Figure 9c.

In this model, the parameters W1, W2, and W3 represent the weight of the network node's exposure to the environment. According to the experimental results, enhancing the contact environment can effectively suppress the spread of rumors by decreasing the effective contact weight (W1) in the process of spreading rumors, increasing the effective contact weight (W2) in the process of dispelling rumors, and replacing the effective contact weight (W3) in the process of rumors. This can prevent the spread of rumors in cyberspace, shorten the duration of rumors, and decrease the scale of rumor dissemination.

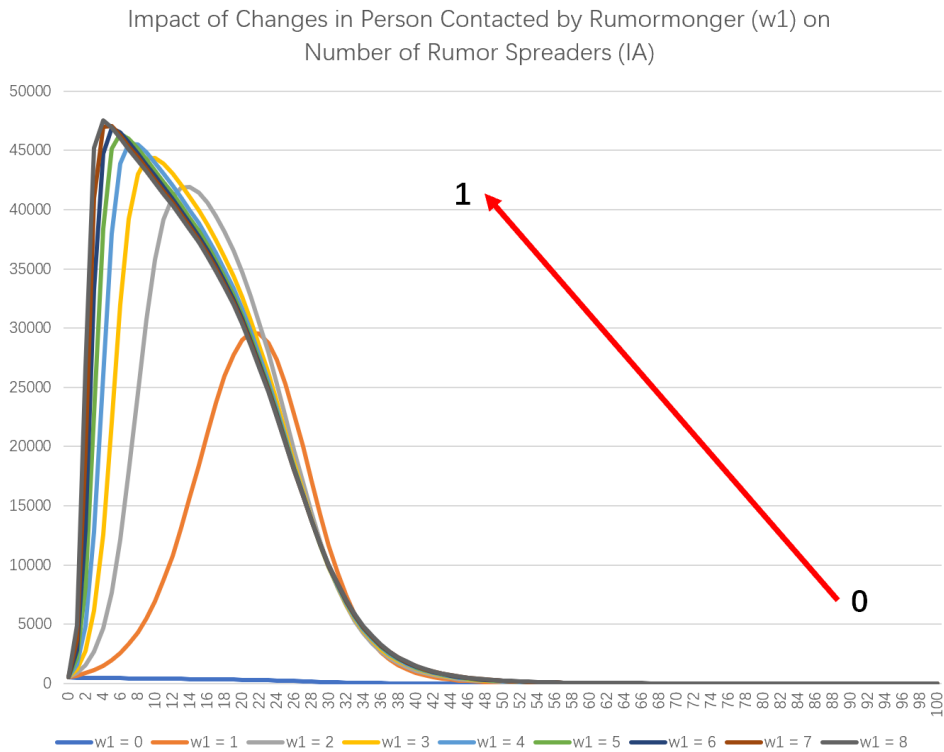

Figure 10a.

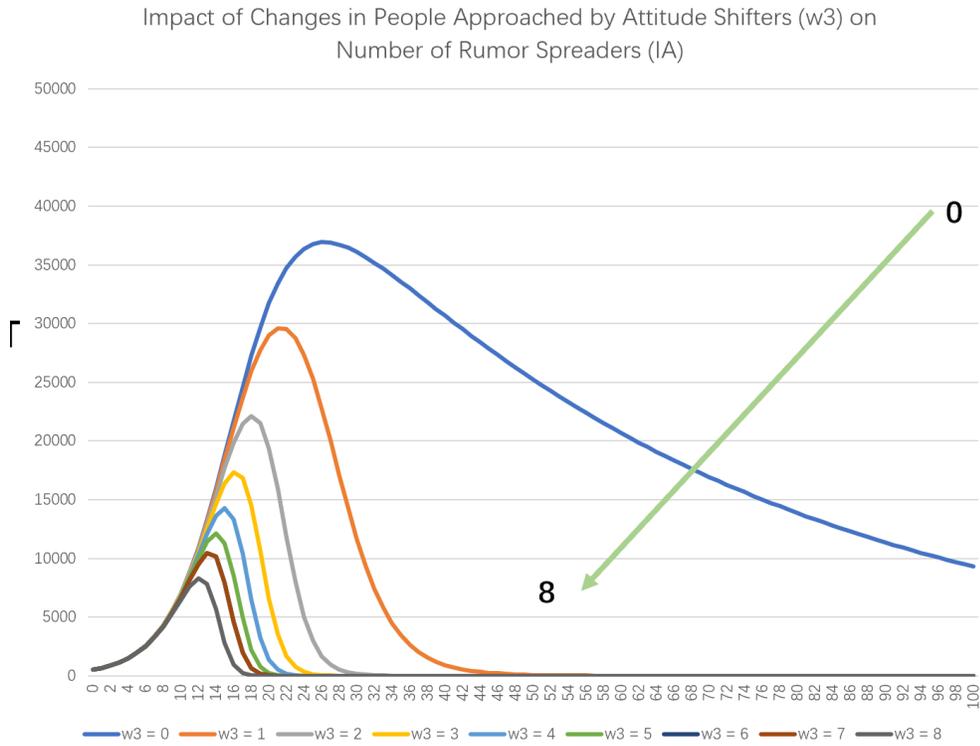

Figure 10b.

When sensitivity analysis is added, it becomes clear that the effective contact weight (W2) in the process of putting an end to rumors is a crucial regulatory factor in the network node's exposure to environmental factors. Flexible control of this parameter can help improve the effect of network bad information governance. These findings from parameter altering experiments highlight the importance of knowledge level and effective contact weight in managing Internet rumors.

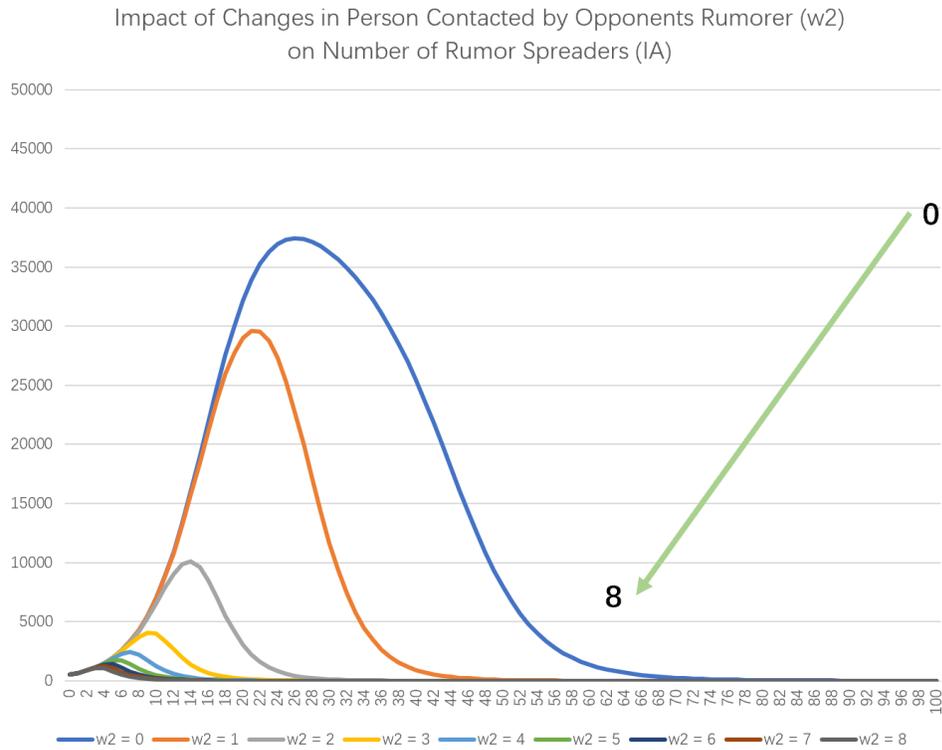

Figure 10c.

**Discussion**

  The experimental findings of this research provide policy options for controlling false information in cyberspace. From the perspective of enhancing the quality of netizens, it is crucial to reduce the rate at which rumors are propagated. Cyberspace is not exempt from the law, and governance entities such as the government and network information service platforms should intensify their enforcement of existing laws to stop the development and spread of rumors. Furthermore, it is important to focus on those who are less able to recognize rumors and are more inclined to spread them, such as children and some senior groups with low levels of knowledge, and take effective steps to improve their capacity to recognize and resist them.

  Additionally, to up the rate of spread of debunking information, the government should intensify the development of rumor-debunking platforms and inspire netizens to actively contribute to the distribution of debunking information. The public should be encouraged to uphold the "clear" ecology of network information content by promptly reporting new online rumors to rumor-reporting platforms. Appropriate rewards should also be given to rumor reporters to create an ecosystem that efficiently manages rumors and cleans up cyberspace.

Moreover, the one-way replacement rate should be boosted. The government and platform management parties should strengthen the public relations and education of netizens. Lifelong quality education for all should also increase netizens' knowledge of their responsibility for maintaining network ethics. Finally, netizens' capacity to make corrections quickly and take part in constructive activities in the event that they misrepresent unfavorable information should be improved to better understand the spread of knowledge.

### EXPERIMENTAL ANALYSIS OF PARAMETERS RELATED TO CITIZENS' SELF-KNOWLEDGE ACCOMPLISHMENT

| PARAMETER REGULATION | MEANING | RUMOR SPREAD SCALE | RUMOR SPREAD PEAK | DURATION OF RUMORS | SENSITIVITY ANALYSIS EFFECTS | GOVERNANCE EFFECT |
|---|---|---|---|---|---|---|
| Other things remain unchanged, gradually increase b1 | The ability of the masses to identify rumors is reduced | Increase | Become higher | Slightly extended | The range of changes in the early stage is large, and soon reaches the state of raging rumors, and the range of changes in the later period is narrowed | Reducing this parameter can achieve the purpose of governance, and the effect will be obvious only when it is reduced to a certain extent |
| Other things remain unchanged, gradually increase b2 | The ability of the public to identify rumors is enhanced | Significantly decreased | Significantly lower | Significantly shortened | The range of rumors caused by each change of parameters is obvious, and the sensitivity is high | Increasing this parameter can achieve the purpose of governance, and the governance effect is the best |
| Other things remain unchanged, gradually increase b3 | The masses actively participate in the rumor dispelling action | Decrease | Become lower | Slightly shortened | The range of changes is gradually slowing down, and the range of rumors in the later period is limited | Increasing this parameter can achieve the purpose of governance, and the early effect is obvious |

Table I

### EXPERIMENTAL ANALYSIS OF CHARACTERISTIC PARAMETERS OF EXPOSURE ENVIRONMENT

| PARAMETER REGULATION | MEANING | RUMOR SPREAD SCALE | RUMOR SPREAD PEAK | DURATION OF RUMORS | SENSITIVITY ANALYSIS EFFECTS | GOVERNANCE EFFECT |
|---|---|---|---|---|---|---|
| Other things remain unchanged, gradually increase w1 | Rumors are rampant | Increase | Become higher | Slightly extended | The range of changes in the early stage is large, and it will soon reach the state of raging rumors, and the range of changes in the later period will shrink | Reducing this parameter can achieve the purpose of governance, and the effect will be obvious only when it is reduced to a certain extent |
| Other things remain unchanged, gradually increase w2 | Rumors are suppressed | Significantly decreased | Significantly lower | Significantly shortened | The variation range of rumors caused by each parameter change is very obvious, and the sensitivity is high | Increasing this parameter can achieve the purpose of governance, and the governance effect is the best |
| Other things remain unchanged, gradually increase w3 | Backpropagation | Decrease | Become lower | Slightly shortened | The range of changes is gradually slowing down, and the range of rumors in the later period is limited | Increasing this parameter can achieve the purpose of governance, and the early effect is obvious |

Table II

**Conclusion**

This research has integrated the foundational components of previous theoretical research on the spread of false information into a competitive dissemination model of rumor events. We have added a one-way state replacement process based on the index provided by a research search engine company to make the model design more practical. By examining the influence mechanism of the competitive communication process and conducting simulation observations on the six regulatory elements, we have drawn insightful results with theoretical and innovative value. The simulation scenario's environment is plausible and closely resembles reality, displaying the impact of various factors on the propagation of rumors simply and intuitively. These findings can be used as practical guidelines for the competitive broadcast of good and bad information on networks for governance, increasing the effectiveness of governance.

However, it is important to note that while our model can describe and explain related problems to some extent, the actual network's scale and structure are enormous and complex. The dynamic mechanism that incorporates numerous aspects of human behavior and decision-making has not yet been perfected in a system. Information may be competitively disseminated many-to-many in the real network environment, with a large magnitude, and the granularity of related parameters may continue to be subdivided into new influencing factors. The mean field has no impact on the local network's characteristics. Therefore, numerous issues still need to be resolved before they can be fully revealed.

In the future, better solutions will be put forth for the design of complex models of bad information governance, and the ongoing development of technologies like information tide sources and information processing will bring about better opportunities for the establishment of a more comprehensive cyberspace governance system of complex networks, resulting in more effective cyberspace governance.